\documentclass[aps,prd,twocolumn,floatfix,preprintnumbers,nofootinbib,superscriptaddress]{revtex4}
\usepackage{mathtools}
\usepackage{graphicx}
\usepackage{color}
\usepackage{natbib}
\usepackage{multirow}
\usepackage{amsmath}
\usepackage[amssymb]{SIunits}

\newcommand{\dd}{\mathrm{d}}
\begin{document}

\title{CMB bounds on disk-accreting massive primordial black holes}
\author{Vivian Poulin}
\author{Pasquale D. Serpico}
\affiliation{LAPTh, Universit\'e Savoie Mont Blanc \& CNRS, 74941 Annecy Cedex, France}
\affiliation{Institute for Theoretical Particle Physics and Cosmology (TTK), RWTH Aachen University, D-52056 Aachen, Germany}
\author{Francesca Calore}
\affiliation{LAPTh, Universit\'e Savoie Mont Blanc \& CNRS, 74941 Annecy Cedex, France}
\author{S\'ebastien Clesse}
\affiliation{Institute for Theoretical Particle Physics and Cosmology (TTK), RWTH Aachen University, D-52056 Aachen, Germany}
\author{Kazunori Kohri}
\affiliation{Theory Center, IPNS, KEK, Tsukuba 305-0801, Ibaraki, Japan}
\affiliation{The Graduate University of Advanced Studies (Sokendai), Tsukuba 305-0801, Ibaraki,Japan}

\date{\today}

\preprint{LAPTH-022/17, TTK-17-22, KEK-Cosmo-207, KEK-TH-1988}

\begin{abstract} 
Stellar-mass Primordial Black Holes (PBH) have been recently reconsidered as a Dark Matter (DM) candidate, after the aLIGO discovery of several binary BH mergers with masses of tens of $M_{\odot}$. Matter accretion on such massive objects leads to the emission of high-energy photons, capable of altering the ionization and thermal history of the universe. This in turn affects the statistical properties of the cosmic microwave background (CMB) anisotropies. Previous analyses have assumed {\it spherical} accretion. We argue that this approximation likely breaks down and that {\em an accretion disk} should form in the dark ages.  Using the most up-to-date tools to compute the energy deposition in the medium, we derive constraints on the fraction of DM in PBH. Provided that disks form early on, even under conservative assumptions for accretion, these constraints exclude a monochromatic distribution of PBH with masses above $\sim 2\, M_\odot$ as the dominant form of DM. The bound on the median PBH mass gets more stringent if a broad, log-normal mass function is considered. A deepened understanding of non-linear clustering properties and BH accretion disk physics  would permit an improved treatment and possibly lead to more stringent constraints. 
\end{abstract}

\maketitle
\section{Introduction}
Despite a wealth of evidences for its existence, the nature of the Dark Matter (DM) composing more than 80\% of the total matter content of our universe remains unknown. 
Particle candidates---e.g. from supersymmetric extensions of the standard model of particle physics---are still the most explored ones, in particular {\em weakly interacting massive particles} (WIMPs),  in which the DM relic density $\Omega_{\rm cdm} h^2= 0.1205$~\cite{Aghanim:2016yuo}, is obtained via the standard freeze-out mechanism.  However, the lack of a WIMP detection via collider, direct, or indirect experiments is now reviving the interest for alternatives. A promising and well-studied {\it macroscopic} alternative to particle DM are primordial black holes (PBH), as recently reviewed in Ref.~\cite{Carr:2016drx}. This scenario has received a lot of attention after the aLIGO discovery of three or four binary black hole (BH) mergers of tens of solar masses~\cite{Abbott:2016blz,Abbott:2016nmj,Abbott:2017vtc}, including one with a progenitor spin misaligned with the orbital momentum. Intriguingly, their merging rate is compatible with the expectation from binaries formed 
in present-day halos by a BH population whose density is comparable to the DM one \cite{Bird:2016dcv,Clesse:2016vqa}, although  Ref.s \cite{Sasaki:2016jop,Raidal:2017mfl} argue that this is significantly  lower than the merger rate of binaries formed in the early universe, which would thus overshoot the aLIGO observed rate.

Black holes in a wide range of masses could have formed in the early universe due to the collapse of $\mathcal{O}(1)$ primordial inhomogeneities \cite{Carr:1974nx,Carr:1975qj,Harada:2013epa},  usually associated to either extended inflationary models (such as hybrid inflation \cite{GarciaBellido:1996qt,2011arXiv1107.1681L,Bugaev:2011wy,Clesse:2015wea}, curvaton scenarios \cite{Kohri:2012yw,Kawasaki:2012wr}, single-field and multi-field models in various frameworks~\cite{Kawasaki:2016pql,Garcia-Bellido:2016dkw,2017arXiv170203901G,Domcke:2017fix,Germani:2017bcs,Ezquiaga:2017fvi,Kannike:2017bxn,Motohashi:2017kbs}), or to first and second-order phase transitions \cite{Jedamzik:1999am,Rubin:2001yw}.
PBH with masses $M\lesssim 10^{-17} M_{\odot}$ evaporate into standard model particles with  a blackbody spectrum (the so-called Hawking radiation \cite{Hawking:1974sw, Hawking:1974rv}), leading to energetic particle injection which can be looked for in cosmic rays \cite{Barrau:2001ev}, $\gamma$ rays \cite{Carr:2016hva} or CMB analysis \cite{2017JCAP...03..043P}. The intermediate mass range up to stellar masses is covered by a number of lensing constraints. From low to high masses, we mention femtolensing in gamma-ray bursts~\cite{2012PhRvD..86d3001B}, microlensing in high-cadence observations of M31~\cite{Niikura:2017zjd} and of the Magellanic clouds~\cite{PalanqueDelabrouille:1997uj,Alcock:2000ph,Tisserand:2006zx}. The latter are however still controversial (e.g. Ref.~\cite{Hawkins:2011qz,Green:2017qoa}), depending on the PBH clustering properties \cite{Clesse:2015wea}; some results even point at a possible detection of anomalous microlensing events \cite{PalanqueDelabrouille:1997uj,Alcock:2000ph}. Additional constraints from neutron stars and white dwarfs in globular clusters also exist in this range~\cite{Capela:2013yf,Capela:2012jz}, but depend on astrophysical assumptions. Stellar mass or heavier PBH are constrained by dynamical properties of ultra-faint dwarf galaxies \cite{Brandt:2016aco,Green:2016xgy,Li:2016utv,Koushiappas:2017chw}, by halo wide binaries \cite{2014ApJ...790..159M}, by X-ray or radio emission~\cite{Gaggero:2016dpq,Inoue:2017csr}, as well as by the cosmic microwave background (CMB) bounds discussed in the following\footnote{Further constraints exist, e.g. based on the emitted gravitational wave background \cite{Nakama:2016gzw,Clesse:2016ajp,Schutz:2016khr,Cholis:2016xvo} or non-gaussianities in the primordial fluctuations \cite{Tada:2015noa,Young:2015kda}, which---while often quite stringent---are model dependent.}. Indeed, due to their gravitational attraction on the surrounding medium, such massive objects accrete matter, which heats up, gets eventually ionized and emits high-energy radiation. In turn, these energetic photons can alter the ionization and thermal history of the universe, affecting the statistical properties of CMB anisotropies. Very stringent constraints (excluding PBH as DM with $M\gtrsim 0.1~M_\odot$) have been thus derived on this scenario already a decade ago~\cite{Ricotti:2007au}. These bounds (as well as their update in Ref.~\cite{Horowitz:2016lib}) have been recently revisited and corrected in Ref.~\cite{Ali-Haimoud:2016mbv} (see also Ref.~\cite{Blum:2016cjs}), yielding significantly weaker constraints $M \lesssim 10-100~M_\odot$ if PBH constitute the totality of the DM, depending on the assumption on radiation feedback.

Although such bounds are usually derived assuming a monochromatic PBH mass function, actual bounds on extended mass functions are typically more stringent~\cite{Green:2016xgy,Kuhnel:2017pwq,Carr:2017jsz}. Also, the time evolution of the initial mass function due to merging events is strongly constrained by purely gravitational CMB bounds: in each merger with comparable BH masses, a few  percent of their mass is converted into gravitational waves, i.e. ``dark'' radiation, a phenomenon that cannot involve more than a small fraction of the DM, due to alterations to the Sachs-Wolfe effect. Essentially no more than one merger per PBH on average is allowed between recombination and now~\cite{Poulin:2016nat}.

In this paper, we revisit the CMB anisotropy constraints on the PBH abundance, which have been derived until now assuming {\it spherical} accretion of matter onto BH. We revisit this hypothesis and find plausible arguments suggesting that an {\em an accretion disk} generically forms in the dark ages, between recombination and reionization possibly already at $z\sim {\cal O}$(1000). A firm proof in that sense would require deeper studies of the non-linear growth of structures at small scales, accounting for the peculiarities of PBH clustering and for the time-dependent building-up of the baryonic component of halos. A first step to motivate such studies, however, is to prove that they have a potentially large impact:  in presence of disks, CMB constraints on PBH improve by (at least) two orders of magnitude, excluding the possibility that  PBH with masses $M \gtrsim 2~M_\odot$ account for the totality of the DM. As we will argue, we expect the bounds to be greatly improved if the baryon velocity at small scales is not coherent and comparable with (or smaller than) their cosmological thermal velocity, and/or if a sizable baryon filling of the PBH halos is present already at $z\gtrsim {\cal O}$(100). 

This article is structured as follows: In Sec.~\ref{sec:essentials}, we  provide a short---and necessarily incomplete---review of the current understanding of accretion,
and discuss its applicability in the cosmological context. The crucial arguments on why we think plausible that the accretion (at least the one relevant for CMB bounds) should proceed via disks is discussed in Sec.~\ref{sec:disks}. In Sec.~\ref{Lumin} we review the expected high-energy luminosity associated to these accretion phenomena and describe benchmark prescriptions used afterwards. Section~\ref{CMBbound} described our procedure on obtaining CMB bounds.  In section \ref{sec:conclu}, we summarize our results and draw our conclusions.

\section{Accretion in cosmology}\label{sec:basics}
\subsection{Essentials on accretion}\label{sec:essentials}

The problem of accretion of a point mass $M$ moving at a constant speed  $v_{\rm rel}$ in a homogeneous gas of number density $n_{\infty}$ (and mass density $\rho_{\infty}$, where the subscript $\infty$ means far away from the point mass)
was first studied by Hoyle and Lyttleton~\cite{1939PCPS...35..405H,1940PCPS...36..424H,1940PCPS...36..325H} in a purely ballistic limit, i.e. accounting only for gravitational effects but no hydrodynamical or thermodynamical considerations. They found the accretion rate (natural units $c=\hbar=k_{\rm B}=1$ are used throughout, unless stated otherwise)
\begin{equation}\label{eq:HLAccRate}
\dot{M}_{\rm HL} \equiv \pi r^2_{\rm HL}\rho_{\infty}v_{\rm rel}\equiv4\pi \rho_{\infty}\frac{(GM)^2}{v_{\rm rel}^3}\,,
\end{equation}
where we introduced the Hoyle-Lyttleton radius $r_{\rm HL}$, the radius of the {\it cylinder} effectively sweeping the medium. This model does not describe the motion of the particles once they reach the  (infinitely thin and dense) accretion line in the wake of the point mass, when pressure and dissipation effects prevail. Also, it is clearly meaningless in the limit of very small velocity $v_{\rm rel}$. A first attempt to address the former problem and account for the accretion column was done by Bondi and Hoyle~\cite{1944MNRAS.104..273B}, suggesting a reduced accretion by up to a factor two. The second problem is linked to neglecting pressure. It has only been solved exactly for an accreting body at rest in a homogeneous gas, when the accretion is spherical by symmetry. Its rate has been computed by Bondi~\cite{Bondi:1952ni}, yielding the so-called {\em Bondi accretion rate}: 
\begin{equation}\label{eq:BondiAccRate}
\dot{M}_{\rm B} \equiv 4\pi \lambda\, \rho_{\infty}c_{{\rm s},\infty}r^2_{\rm B}\equiv 4\pi \lambda \,\rho_{\infty}\frac{(GM)^2}{c_{{\rm s},\infty}^3}\,,
\end{equation}
where  $r_{\rm B}$ is the Bondi radius, i.e. the radius of the equivalent accreting {\it sphere} (as opposed to a cylinder, hence the $4\pi$ geometric factor), $c_{s,\infty}$ is the sound speed far away from the point mass, depending on the pressure ${\rm P}_\infty$ and density $\rho_\infty$, and $\lambda$ is a parameter that describes the deviation of the accretion from the Bondi idealised regime.
 In the cosmological plasma, one typically has:
\begin{eqnarray}
&&c_{{\rm s},\infty}  =  \sqrt{\frac{\gamma{\rm P}_{\infty}}{\rho_{\infty}}}= \sqrt{\frac{\gamma(1+x_{\rm e})T}{m_{\rm p}}}\simeq 6 \frac{\rm km}{\rm s}\sqrt{\frac{1+z}{1000}}\,,\label{cs}\\
&&\Rightarrow r_{\rm B} \equiv \frac{GM}{c_{{\rm s},\infty}^2}  \simeq   1.2 \times 10^{-4} {\rm pc} \frac{M}{M_\odot}\frac{10^3}{1+z}\,,\label{rBnum}
\end{eqnarray}
 $m_{\rm p}$ being the proton mass, and $\gamma$ is the  polytropic equation of state coefficient for monoatomic ideal gas. The approximation at the RHS of Eq.~(\ref{cs}) typically holds for $100\lesssim z\lesssim 1000$. The mean cosmic gas density in the early universe is given by:
\begin{equation}\label{eq:n_gas}
n_{\infty} \simeq \frac{\rho_\infty }{m_{\rm p}} \simeq 200 \,{\rm cm}^{-3}\bigg(\frac{1+z}{1000}\bigg)^3\,.
\end{equation}
Finally, $\lambda$ is a numerical parameter which quantifies non-gravitational forces (pressure, viscosity, radiation feedbacks, etc.) partially counteracting the gravitational attraction of the object.  Historically, Bondi computed the maximal value of $\lambda$ as a function of the equation of state of the gas, finding $\lambda\sim{\cal O}$(1), ranging from 0.25 ($\gamma = 5/3$, adiabatic case) to 1.12 ($\gamma =1$, isothermal case).  

There is no exact computation of the accretion rate accounting for the finite sound speed and a displacement of the accreting object. However, as argued by Bondi in Ref.~\cite{Bondi:1952ni}, a reasonable proxy can be obtained by the quadratic sum of the relative velocity and the sound speed at infinity, which leads to an effective velocity  $v_{\rm eff}^2 = c_{s,\infty}^2+v_{\rm rel}^2$.  We thus define the Hoyle-Bondi radius and rate\footnote{Actually, our rate definition is a factor 2 larger than the original proposal, but has been confirmed as more appropriate even with numerical simulations, see Ref.~\cite{1985MNRAS.217..367S}.}
\begin{equation}\label{eq:HBAccRate}
\dot{M}_{\rm HB} \equiv 4\pi\lambda\,  \rho_{\infty}v_{\rm eff}r^2_{\rm HB}\equiv 4\pi\lambda\,  \rho_{\infty}\frac{(GM)^2}{v_{\rm eff}^3}\,.
\end{equation}

Despite the fact that the Bondi analysis was originally limited to spherical accretion, this formalism is commonly used to treat non-spherical cases, with e.g. formation of an accretion disk, by choosing an appropriate value for $\lambda$. Although it has been shown for instance that the simple analytical formulae can overestimate accretion in presence of vorticity~\cite{Krumholz:2004vj} or underestimates it in presence of  turbulence~\cite{Krumholz:2005pb}, typically Eq.~(\ref{eq:HBAccRate}) provides a reasonable order-of-magnitude description of the simulations (see for instance~\cite{Mellah:2015sja} for a recent simulation and interpolation formulae).

\subsection{Relative baryon-PBH velocity and disk accretion in the early universe}\label{sec:disks}
 In the cosmological context, one might naively estimate the relative velocity between DM and baryons to be of the order of the thermal baryon velocity or of the speed of sound, Eq.~(\ref{cs}).  In that case, the appropriate accretion rate would be the Bondi one, Eq.~(\ref{eq:BondiAccRate}). 
 The situation is however more complicated, since at the time of recombination the sound velocity drops abruptly and the baryons, which were initially tightly coupled to the photons in a standing acoustic wave, acquire what is an eventually supersonic relative stream with respect to DM, coherent over tens of Mpc scales. 
In linear theory, one finds that  the square root of the variance  of the relative baryon-DM velocity is basically constant before recombination and then drops linearly with $z$~\cite{Tseliakhovich:2010bj,Dvorkin:2013cea}:
\begin{equation}
\sqrt{\langle v_{\rm L}^2\rangle}\simeq {\rm min} \left[1, \frac{1+z}{1000}\right]\times 30\, {\rm km/s}\,.\label{vbulk}
 \end{equation}
 Yet, this is a linear theory result, and it is unclear if it can shed any light on the accretion, which depends on very small, sub-pc scales (Bondi radius, see Eq.~(\ref{rBnum})). 
 In Ref.~\cite{Tseliakhovich:2010bj}, the authors first studied the problem of small-scale perturbation growth into such a configuration, by a perturbative expansion of the fluid equations for DM, baryons, and the Poisson equation around the exact solution with uniform bulk motion given by Eq.~(\ref{vbulk}), {\it further assuming zero density contrast, and zero Poisson potential.}
Their results suggest that  small-scale structure formation and the baryon settling into DM potential wells is significantly delayed with respect to simple expectations. Equation~(\ref{vbulk}) has also entered recent treatments of the Hoyle-Bondi PBH accretion rate, see Ref.~\cite{Ali-Haimoud:2016mbv}, yielding a correspondingly suppressed accretion. In particular, by taking the appropriate moment of the function of velocity entering the luminosity of accreting BH over the velocity distribution,  Ref.~\cite{Ali-Haimoud:2016mbv} found
\begin{equation}
v_{\rm eff}\equiv\left\langle\frac{1}{ (c_{s,\infty}^2+v_{\rm L}^2)^3 }\right\rangle^{-1/6}\simeq \sqrt{c_{s,\infty}\sqrt{\langle v_{\rm L}^2\rangle}}\,,\label{veff}
 \end{equation}
with the last approximation only valid if $c_{s,\infty}\ll\sqrt{\langle v_{\rm L}^2\rangle}$, which is acceptable at early epochs after recombination, of major interest in the following.

The application of the above perturbative (but non-linear) theory to the relative motion between PBH and the baryon fluid down to sub-pc scales appears problematic. A first consideration is that the behavior of an ensemble of PBH of stellar masses is very different from the ``fluid-like'' behavior adopted for microscopic DM candidates like WIMPs. The discreteness of PBHs is associated to a ``Poissonian noise'', enhancing the DM power spectrum at small scale, down to the horizon formation one~\cite{Afshordi:2003zb,Chisholm:2005vm,Zurek:2007gn,Gong:2017sie}.  Our own computation suggests that a density contrast of $\mathcal{O}(1)$ is attained at $z\simeq 1000$ at a comoving scale as large as $k_{\rm NL}\sim 10^3$ Mpc$^{-1}$ for a population of 1$\,M_\odot$ PBH whose number density is comparable to the DM one. Even allowing for fudge factors (e.g. $f_{\rm PBH}\sim 0.1$, different mass) the non-linearity scale is unavoidably pertinent to the scales of interest. In fact, the PBH formation mechanism itself {\em is} a non-linear phenomenon, and peaks theory suggests that PBH are likely {\it already born} in clusters, on the verge of forming  bound systems~\cite{Chisholm:2005vm,Chisholm:2011kn}.  Our first conclusion is that the application of the scenario considered in Refs.~\cite{Tseliakhovich:2010bj,Dvorkin:2013cea} to the PBH case is not at all straightforward. In particular, a more meaningful background solution around which to perturb would be the one of vanishing initial baryon perturbations in the presence of an already formed halo (and corresponding gravitational potential) at a scale $k_{\rm NL}\gtrsim 10^3$ Mpc$^{-1}$. 
A second caveat is that the treatment in Refs.~\cite{Tseliakhovich:2010bj,Dvorkin:2013cea} uses a fluid approximation, i.e. it does not account for ``kinetic'' effects such as the random (thermal) velocity distribution around the bulk motion velocity given by Eq.~(\ref{vbulk}). One expects that ``cold'' baryons (statistically colder than the average) would already settle in the existing PBH halo at early time, forming a virialized system---albeit still under-dense in baryons, with respect to the cosmological baryon to DM ratio. 
One may also worry about other effects, such as shocks and instabilities, which may hamper the applicability of the approach of Ref.~\cite{Tseliakhovich:2010bj} to too small scales and too long times. 

Assuming that the overall picture remains nevertheless correct in a more realistic treatment, we expect that the PBH can generically accrete from two components: the high-velocity, free-streaming fraction at cosmological density and diminished rate of Eqs.~(\ref{eq:HBAccRate}) and (\ref{veff}), as considered in Ref.~\cite{Ali-Haimoud:2016mbv}, and a virialized component, of initial negligible density but growing with time and eventually dominating, with typical relative velocity of the order of the virial ones. If we normalize to the Milky Way halo ($10^{12}\,M_\odot$) value $v_{\rm vir}\sim 10^{-3}\,c$, and adopt the simple scaling of the velocity with the halo mass over size,  $v_{\rm vir}(M_{\rm halo})\propto (M_{\rm halo}/d_{\rm halo})^{1/2}\propto M_{\rm halo}^{1/3}$, we estimate  $v_{\rm vir}\sim 0.3\,$km/s to $3\,$km/s for a halo mass of $10^{3}\,M_\odot$ to $10^{6}\,M_\odot$. The latter  roughly corresponds  to the smallest dwarf galaxies one is aware of, see e.g.~\cite{Bonnivard:2015xpq}~\footnote{The PBH distribution can hardly be dominated by heavier clumps, or the lack of predicted structures at the dwarf scales would automatically exclude them as dominant DM component.}. At $z\simeq {\cal O}(1000)$, it is likely that the fast, unbound baryons constitute the dominating source of accretion. But at latest when the density of the virialized baryon component attains values comparable to the {\it cosmological average} density---which given the $z$-dependences Eq.~(\ref{cs}) and Eq.~(\ref{vbulk}) appears unavoidable for $z\lesssim {\mathcal O}(100)$---the accretion is dominated by this halo-bound component. 

After these preliminary considerations, we are ready to discuss disk formation. The basic criterion used to assess if a disk forms is to estimate the angular momentum of the material at the accretion distance: if this is sufficient to keep the matter in Keplerian rotation at a distance $r_{\rm D}\gg 3\, r_{\rm S}$ (i.e. well beyond the innermost stable orbit, where we introduced the Schwarzschild radius $r_{\rm S}\equiv 2\,G\,M$) at least for BH luminosity purposes, dominated by the region close to the BH, a disk will form~\cite{1976ApJ...204..555S,1977ApJ...216..578I, Ruffert:1999ch, 2002MNRAS.334..553A}.  To build up angular momentum, the material accreted at the Hoyle-Bondi distance along different directions must have appreciable velocity or density differences. The angular momentum per unit mass of the accreted gas scales like
\begin{equation}
l\simeq\left(\frac{\delta\rho}{\rho}+\frac{\delta v}{v_{\rm eff}}\right)v_{\rm eff}r_{\rm HB}\,,\label{lscale}
\end{equation}
where $\delta\rho/\rho$ represent typical inhomogeneities at the scale $r_{\rm HB}$ in the direction {\it orthogonal} to the relative motion PBH-baryons, and $\delta v/v_{\rm eff}$
the analogous typical velocity gradient at the same scale (see e.g.~\cite{2002MNRAS.334..553A}). 
The above quantity can be compared to the specific angular momentum of a Keplerian orbit,
\begin{equation}
l_{\rm D} \simeq r_{\rm D}v_{\rm Kep}(r_{\rm D}) \simeq  \sqrt{GMr_{\rm D}}\,,\label{kepdisk}
 \end{equation}
to extract $r_D$. For instance, in the case of inhomogeneities,   if we adopt the effective velocity at the RHS of Eq.~(\ref{veff}) as a benchmark, as in Ref.~\cite{Ali-Haimoud:2016mbv}, we obtain:
\begin{equation}
\frac{r_D}{r_S}\simeq \left(\frac{\delta \rho}{\rho}\right)^2 \frac{c^2}{2\,v_{\rm eff}^2}\simeq 2.5\times10^8\left(\frac{\delta \rho}{\rho}\right)^2\left(\frac{1000}{1+z}\,\right)^{3/2}\,,\label{deltarhocond}
\end{equation}
so that, already soon after recombination, gradients $\delta \rho/\rho \gg 10^{-4}$ in the baryon flow on the scale of the Bondi radius are sufficient for a disk to form. We find this to be largely satisfied already at $z\sim 1000$ because of the ``granular'' potential due to neighboring PBHs.

Equivalently, given the similar way the fractional fluctuation of velocity and density enter Eq.~(\ref{lscale}), the condition for a disk to form can be written as a lower limit on the absolute value of the velocity perturbation amounting to
\begin{equation}
\delta v\gg 1.5\, \left(\frac{1+z}{1000}\,\right)^{3/2}\,{\rm m/s}\,.\label{diskcriterion3}
 \end{equation}
At least  the component of virialized baryons, whose velocity dispersion is $\gtrsim 0.1$ km/s as argued above, should easily match this criterion.

But even for a ``ideal'',  free-streaming homogeneous gas moving at a bulk motion comparable to Eq.~(\ref{vbulk}) without any velocity dispersion, the disk formation criterion is likely satisfied, if the 
non-linear PBH motions at small scales are taken into account. Since this is in general a complicated problem, we cannot provide a cogent proof, but the following argument makes us confident that this is a likely circumstance. In general, the BH motion within its halo at very small scale is influenced by its nearest neighbors. The simplest scenario (see for instance~\cite{Sasaki:2016jop}) amenable to analytical estimates is that a sizable fraction of PBH forms binary systems with their nearest partner, under the tidal effect of the next-to-nearest. 
According to~\cite{Sasaki:2016jop}, for PBH constituting a sizable fraction of the DM, it is enough for their distance to be only slightly below the average distance at matter-radiation equality for a binary to form. Under the assumption of an isotropic PBH distribution and monochromatic PBH mass function of mass $M$, this distance can be estimated as
\begin{equation}
d \sim \left(\frac{3M}{4\pi\rho_{\rm PBH}}\right)^{1/3}=\frac{1}{1+z_{\rm eq}}\left(\frac{2G M}{H_0^2 f_{\rm PBH}\Omega_{\rm DM}}\right)^{1/3}\,,
\end{equation}
i.e.
\begin{equation}
d\sim 0.05\,{\rm pc}\left(\frac{M}{f_{\rm PBH}\,M_\odot}\right)^{1/3}\frac{3400}{1+z_{\rm eq}}\,.\label{dfirst}
\end{equation}
If bound, the two PBH (each of mass $M$) orbit  around the common center of mass on an elliptical orbit whose major semi-axis is $a$  with the Keplerian angular velocity 
\begin{equation}
\omega =  \sqrt{\frac{2\,G\,M}{a^3}}\,.\label{omegaorbit}
\end{equation}
We conservatively assume $a= d/2$ for a quasi-circular orbit, although for 
the very elongated orbits usually predicted for PBH a value $a=d/4$ is closer to reality. Note that the orbital size of the order of Eq.~(\ref{dfirst}) is typically larger than (or at most comparable to) the Bondi-Hoyle radius, so that to a good approximation  the gas---assumed to have a bulk motion with respect to the PBH pair center of mass---accretes around a single PBH, which is however rotating with respect to it. 

In the PBH rest-frame, Eq.~(\ref{lscale}) is simply replaced by  
\begin{equation}
 l\simeq  \omega\,r_{\rm {HB}}^2\,,
 \end{equation}
 or, equivalently, one can apply Eq.~(\ref{diskcriterion3}) with $\delta v= \omega\,r_{\rm HB}$.

If we adopt the effective velocity at the RHS of Eq.~(\ref{veff}), this leads to  the disk formation condition ($z\lesssim 1000$):
 \begin{equation}
  f_{\rm PBH}^{1/2}\frac{M}{M_\odot}\gg  \left(\frac{1+z}{730}\right)^{3} \,.
 \end{equation}
Whenever $M\gtrsim M_\odot$ and PBH constitute a sizable fraction of the DM, this is satisfied at the epoch of interest for CMB bounds.

In fact,  it has been shown in Ref.~\cite{2017JCAP...03..043P,Slatyer:2016qyl} that most of the constraining power of CMB anisotropies on exotic energy injection {\em does not} come from redshift 1000 and above, rather around a typical redshift of $\sim 300$ for an energy injection rate scaling like $\propto(1+z)^3$. In the problem at hand, the constraining power should be further skewed towards  lower redshifts, given the growth of the signal at smaller $z$ due to the virializing component.

We believe that these examples show that disk formation at relatively early times after recombination is a rather plausible scenario, with spherical accretion which would rather require physical justification. 
Note that we have improved upon the earlier discussion of this point in Ref.~\cite{Ricotti:2007au} by taking into account the essential ingredient that stellar mass PBH are clustered in non-linear structures at small scales and early times, greatly differing from WIMPs in that respect.  
In the following, we shall assume that the disk forms at all relevant epochs for setting CMB bounds, and deduce the consequences of this Ansatz. In the conclusions, we will comment on the margins for improvements
over the current treatment.

\subsection{Luminosity}\label{Lumin}
In addition to $\dot{M}$, the second crucial quantity for accretion luminosity is the radiative efficiency factor $\epsilon$, which simply relates the {\em accretion luminosity} $L_{\rm acc}$ to the accretion rate in the following way: 
\begin{equation}
L_{\rm acc}=\epsilon\dot{M}.
\end{equation}
The radiative efficiency is itself tightly correlated with the accretion geometry and thus the accretion rate, since it directly depends on the temperature, density and optical thickness of the accretion region. Hence, a coherent analysis determines both parameters $\lambda$ and $\epsilon$ jointly. In practice, no complete, first-principle theory exists, although a number of models have been developed to compute $L_{\rm acc}$ (which is the main observable in BH physics) under different assumptions and approximations. A typical fiducial value is $\epsilon=0.1$, to be justified below. A useful benchmark upper limit to $L_{\rm acc}$ 
is the so-called Eddington luminosity, $L_E=4\pi G M m_p/\sigma_T=1.26\times 10^{38}\,(M/M_\odot)\,$erg/s, which is the luminosity at which electromagnetic radiation pressure (entering via the Thomson cross section $\sigma_T$) balances the inward gravitational force in a hydrogen gas, preventing larger accretion, unless special conditions are realized. In practice, for the parameters of cosmological interest, it turns out that we will always be below $L_E$.

The simplest and most complete theoretical treatment applies to spherical accretion, going back to Shapiro in Refs.~\cite{1973ApJ...180..531S,1973ApJ...185...69S} in the case of non-rotating BHs and Ref.~\cite{1974ApJ...189..343S} for rotating (Kerr) BHs, accounting for relativistic effects. Since we have argued that this case is unlikely to apply to the cosmological context of interest, we will not review it here, but address for instance to Ref.~\cite{Ali-Haimoud:2016mbv} for a recent and detailed treatment. We will only refer to this case for comparison purposes, and for these cases we follow the equations in Ref.~\cite{Ali-Haimoud:2016mbv}.

For {\it moderate or low disk} accretion rate, which is the case of interest here, there are two main models:

If the radiative cooling of the gas is {\it efficient}, a geometrically thin disk forms, which radiates very efficiently. This is the ``classical'' disk solution obtained almost half a century ago by  Shakura and Sunyaev~\cite{Shakura:1972te}. In this case, the maximal energy per unit mass available is uniquely determined by the binding energy at the innermost stable orbit. This can be computed accurately in General Relativity, yielding $\epsilon$ from 0.06 to 0.4 when going from a Schwarzschild to a maximally rotating Kerr BH. This range, which justifies the benchmark value $\epsilon=0.1$ mentioned above, is often an upper limit to the radiative efficiency actually inferred from BH observations. Also note that, since the disk can efficiently emit radiation, the temperatures characterizing the disk emission are relatively low, below a few hundreds of keV. 

If the radiative cooling of the gas is {\it inefficient}, then hot and thick/inflated disks (or torii) form, with advection and/or convective motions dominating the gas dynamics and inefficient equilibration of ion and electron temperature, with the former that is much higher and can easily reach tens of MeV. This regime is widely (albeit with a little abuse of notation) known under the acronym ADAF, ``advection-dominated accretion flow'' (see~\cite{Yuan:2014gma} for a review).
It has been discovered in the pioneering articles~\cite{Ichimaru:1977uf} and later \cite{Rees:1982pe}, but has been extensively studied only after its ``rediscovery'' and  1D self-similar analytical treatment in Ref.~\cite{Narayan:1994xi}. 
It is worth noting that in the ADAF solution, the viscosity $\alpha$ plays a fundamental role in accretion: 
Indeed the viscously liberated energy is not radiated and dissipated away, but instead is conveyed into the optically thick gas towards the center. 
As a consequence, the accretion rate is typically diminished by an order of magnitude with respect to the Bondi rate with $\lambda =1$ (see~\cite{Narayan:2002ss} for a short pedagogical overview). In practice, $\alpha$ is degenerate with the previously introduced parameter $\lambda$, so that one might roughly capture this effect by assuming as benchmark $\lambda=0.1$. In ``classical'' ADAF models, the efficiency scales roughly linearly with $\dot{M}$, attaining (and stabilizing at) a value of the order of 0.1 only for a critical accretion which is about $0.1\,L_E$. Overall, this class of models provides a moderately satisfactory description (at least for $\alpha\lesssim 0.1$) of ``median'' X-ray observations of nuclear regions of supermassive black holes, see e.g.~\cite{Pellegrini:2005pi} (in particular the lower dashed curve in Fig. 3). 
\begin{figure}
\centering
\includegraphics[scale=0.38]{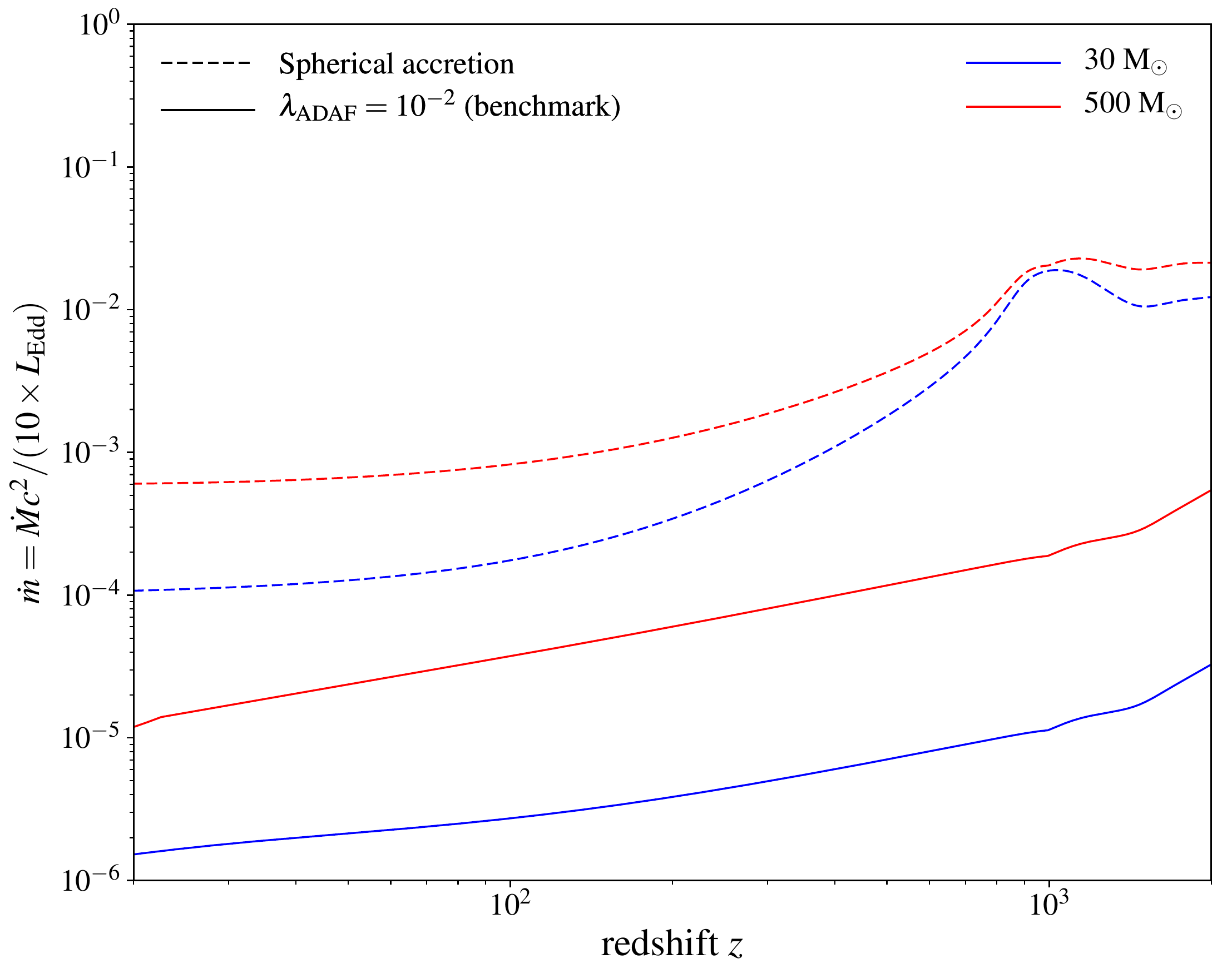}
\includegraphics[scale=0.38]{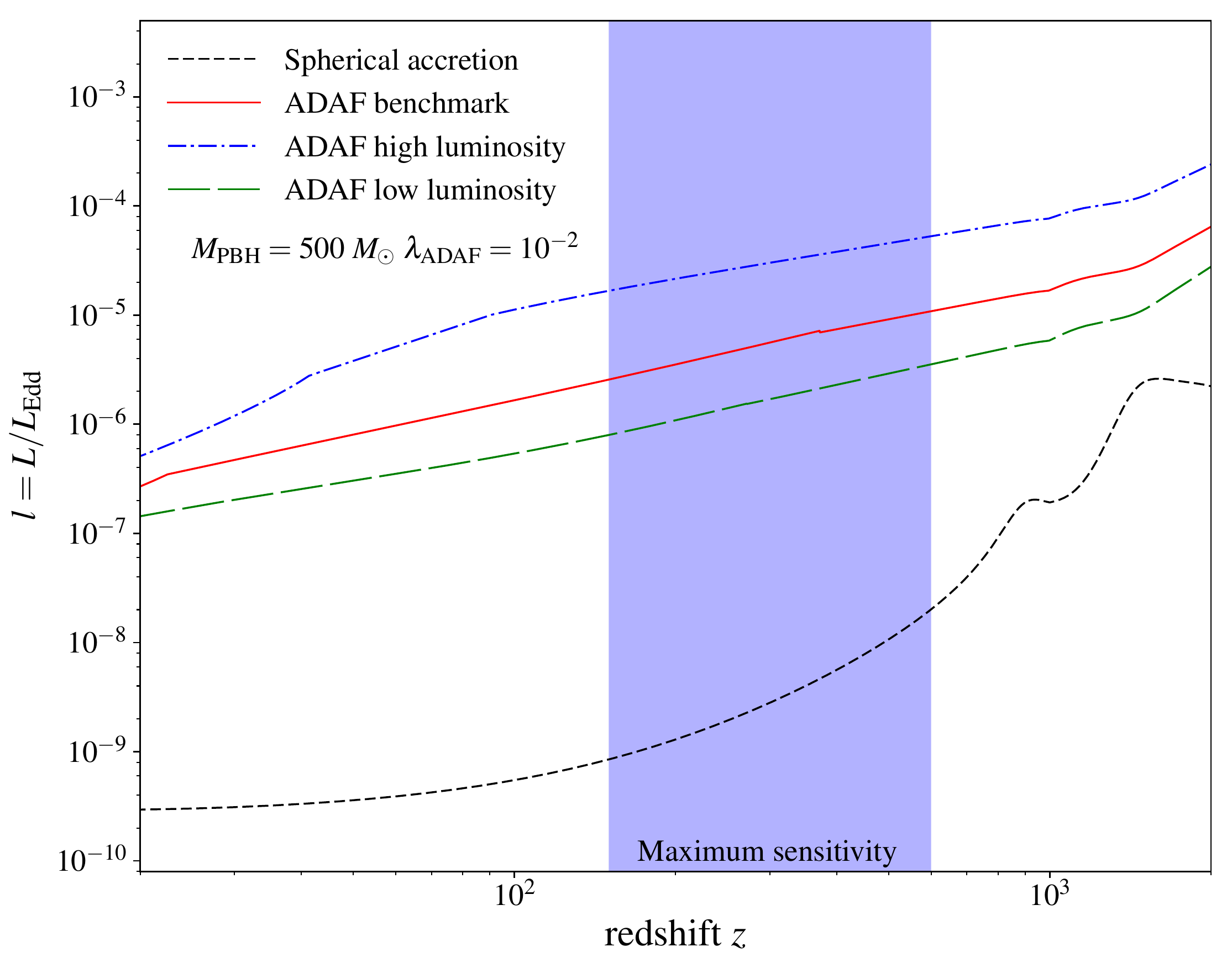}
\caption{{\em Top panel:} The dimensionless accretion rate $\dot{m}$ as a function of redshift for different accretion modeling and PBH mass. Our benchmark model corresponds to the result of simulations attested by observations. {\em Bottom panel:} The dimensionless luminosity $l$ as a function of redshift for different accretion modeling. The benchmark model stands for $\delta = 0.1$, while the low-luminosity and high-luminosity scenarii corresponds to $\delta = 10^{-3}$ and $0.5$ respectively. \label{fig:PBHAccretion}}
\end{figure}

A further refinement takes into account that gas outflows and jets typically accompany this regime, so that the accretion rate becomes in general a function of radius~\cite{Blandford:1998qn}. We will still normalize the (diminished) accretion rate responsible for the bulk of the luminosity to the one at the Bondi radius.
For a specific example, we rely on some recent numerical solutions~\cite{2012MNRAS.427.1580X} which suggest: i) On the one hand, a more significant role of outflows, so that only $\sim 1\%$ of the accretion rate at the Bondi radius is ultimately accreted in the inner region most relevant for the luminosity of the disk. We shall model that by benchmarking $\lambda=0.01$. ii) On the other hand, an increase of the
fraction, $\delta$, of the ion energy shared by electrons. Typically, in classical ADAF models, such a fraction is considered to be very small, $\delta \ll 1$.
A greater efficiency $\delta$ implies a corresponding higher efficiency $\epsilon$, somewhat intermediate between the thin disk and the classical ADAF solution, also scaling with a milder power of the mass accretion ($\epsilon \propto \dot{M}^{0.7}$) at low accretion rates. In Ref.~\cite{2012MNRAS.427.1580X}, suitable fitting formulae have been provided, which we rely upon in the following. In particular, we adopt the parameterization in Eq.~(11), with parameters taken from Tab.~1 for the ADAF accretion rate regime.
In Fig.~\ref{fig:PBHAccretion}, we compare the spherical case with $v_{\rm eff} = \sqrt{c_{s,\infty} \langle v_L\rangle^{1/2}}$ to our benchmark $\delta = 0.1$, as well as a more optimistic $\delta = 0.5$ and a more pessimistic\footnote{It is worth noting that such a low value is reported in Ref.~\cite{2012MNRAS.427.1580X} rather for historical reasons, being associated to the early analytical solutions of Ref.~\cite{Narayan:1994xi} and thus being an old benchmark, than because of theoretical or observational arguments related e.g. to Sgr $A^*$: The authors of Ref.~\cite{2012MNRAS.427.1580X} make clear that all evidence points to a higher range for $\delta$, with $\delta=0.1$ being on the {\it conservative} side, and any $\delta \lesssim 0.3$ is in agreement with data from Sgr $A^*$~\cite{Narayan:2002ss}.} $\delta = 10^{-3}$: the accretion rate (top panel) reduces when a disk forms (independently of $\delta$), but the luminosity (bottom panel) is enhanced. Since in the redshift range of interest (blue band in bottom panel of Fig.~\ref{fig:PBHAccretion}, according to~\cite{2017JCAP...03..043P,Slatyer:2016qyl}) the latter is enhanced despite the fact that the former is reduced (whatever the value of $\delta$), we expect the CMB bound to improve appreciably in our more realistic disk accretion scenario.

\section{Computing the CMB bound}\label{CMBbound}
The total energy injection rate per unit volume  is:
\begin{equation}\label{eq:dQdt}
\frac{\dd E}{\dd V \dd t}=L_{\rm acc}n_{\rm pbh}=L_{\rm acc}f_{\rm pbh}\frac{\rho_{\rm DM}}{M}\,.
\end{equation}
However, not all radiation is equally effective: to compute the impact on the CMB we need to  quantify what amount of this injected energy is deposited into the medium, either through heating, ionization or excitation of the atoms. The modifications of the free electron fraction $x_{\rm e}$ are eventually responsible for the CMB bound. For a given energy differential luminosity spectrum $L_\omega$, the key information is encoded in the {\em energy deposition functions per channel} $f_c(z,x_{\rm e})$ by means of a convolution with the transfer functions $T_c(z',z,E)$ (which we take from Ref. \cite{Slatyer15-2}) according to:
\begin{eqnarray}
f_c(z,x_{\rm e}) & \equiv & \frac{\dd E/(\dd V \dd t)\big|_{{\rm dep},c}}{\dd E/(\dd V\dd t)\big|_{\rm inj}} \label{fzexpr}\\
& = & H(z)\frac{\int \frac{\dd \ln(1+z')}{H(z')}\int T(z',z,\omega) L_\omega \dd \omega}
{\int L_\omega \dd \omega}\,.\nonumber
\end{eqnarray}
The only ingredient left is thus the spectrum of the radiation emitted via BH accretion. Note that it is only the shape that enters Eq.~(\ref{fzexpr}), which is indeed an efficiency function, while the overall normalization was discussed in Sec.~\ref{Lumin}.  In the spherical accretion scenario (see~\cite{1973ApJ...180..531S,1973ApJ...185...69S,Ali-Haimoud:2016mbv}) the spectrum is dominated by Bremsstrahlung emission, with a mildly decreasing frequency dependence over several decades and a cutoff given by the temperature of the medium near the Schwarzschild radius $T_s$ 
\begin{equation}\label{eq:L_nu}
L_{\omega}\propto \omega^{-a}\exp(-\omega/T_s)\,,
\end{equation}
where $T_s\sim {\cal O}(m_{\rm e})$ (we used 200 keV in the following for definiteness) and $|a|\lesssim 0.5$ ($a=0$ was used in ~\cite{Ali-Haimoud:2016mbv}).

\begin{figure}
\centering
\includegraphics[scale=0.38]{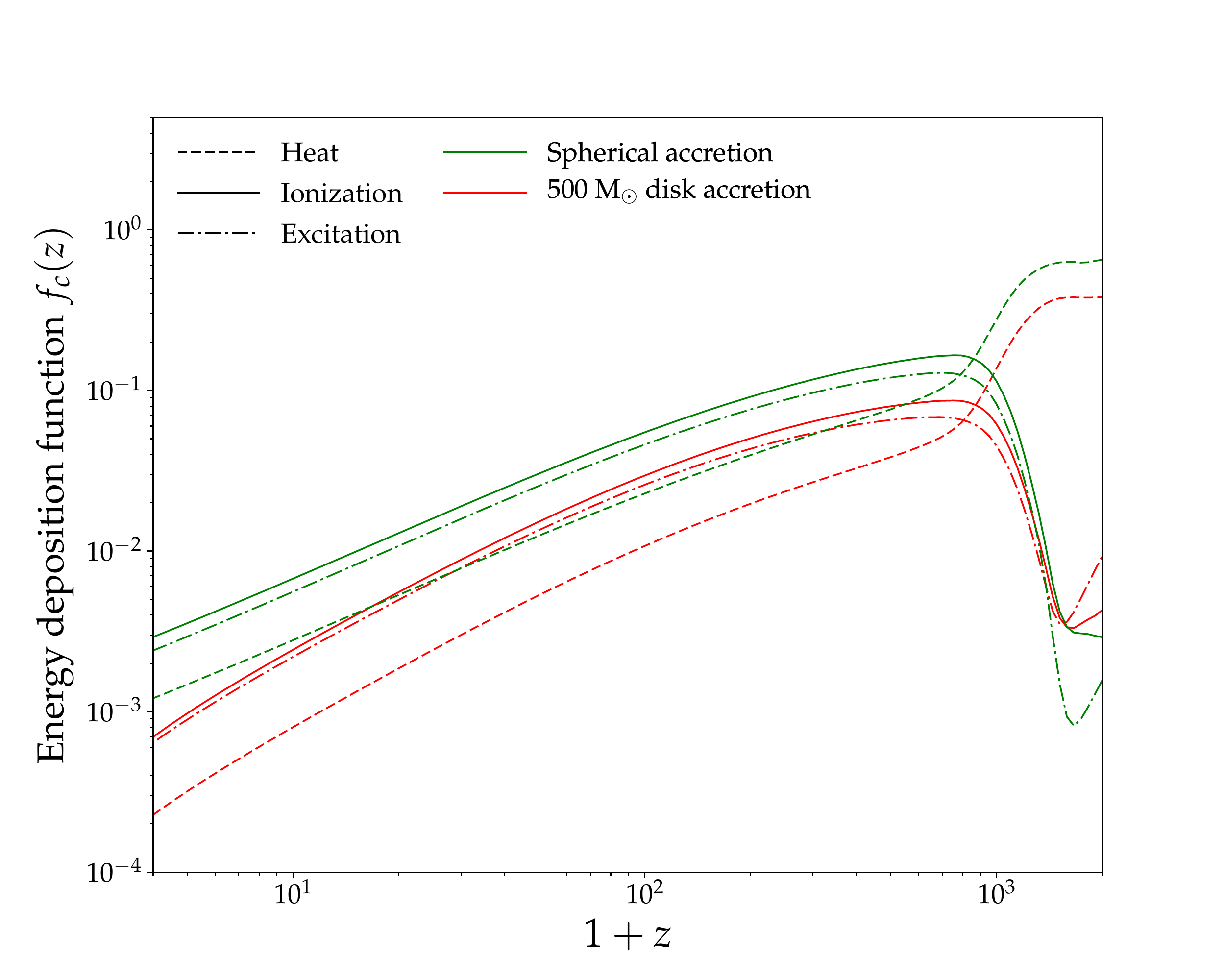}
\includegraphics[scale=0.38]{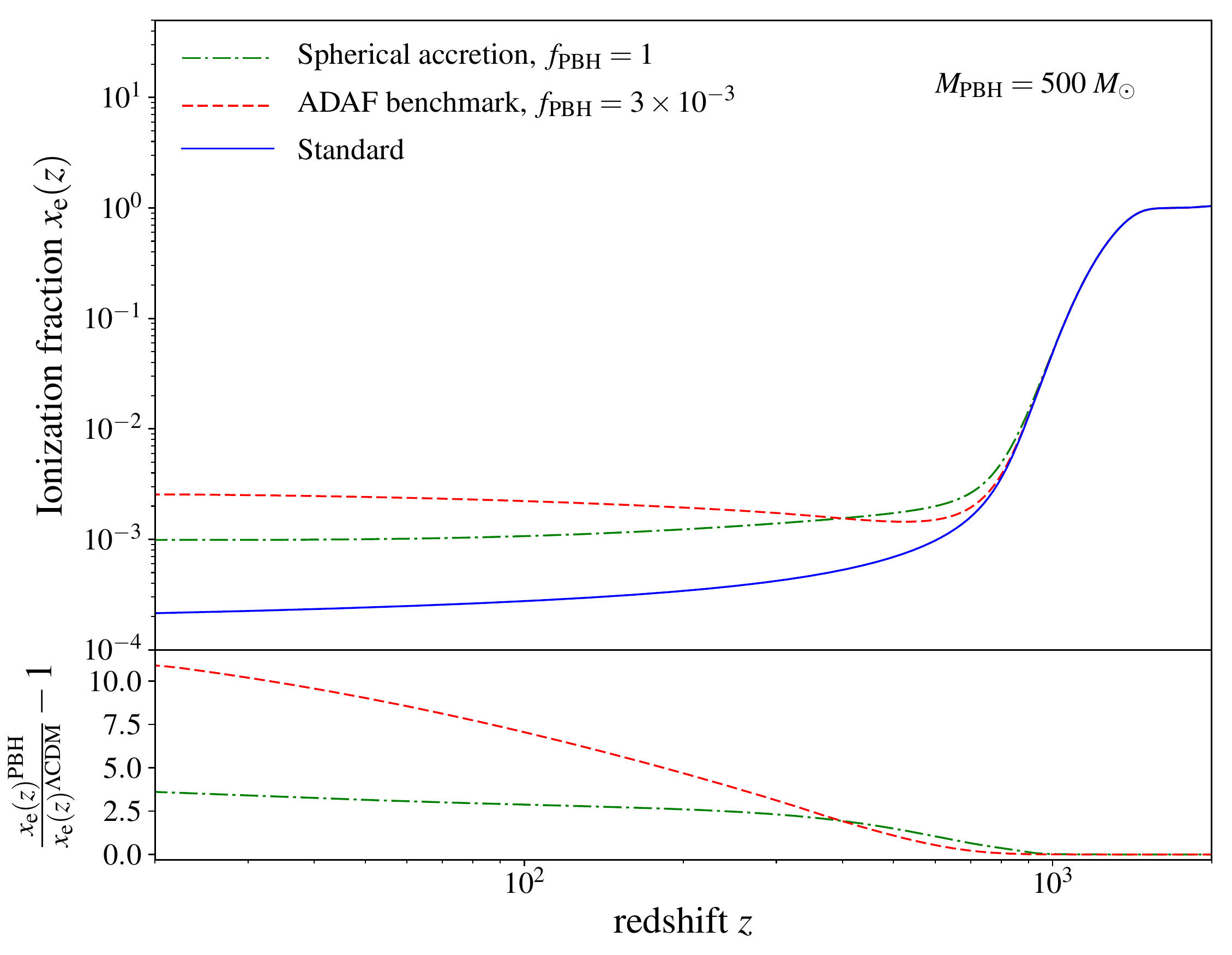}
\caption{{\em Top panel:} Energy deposition functions computed following ref.~\cite{Slatyer15-2} in the case of accreting PBH. {\em Bottom panel:} Comparison of the free electron fractions obtained for a monochromatic population of PBH with masses 500 $\,M_\odot$ depending on the accretion recipe used. The curve labelled ``standard'' refers to the prediction in a $\Lambda$CDM model whose parameters have been set to the best fit of Planck 2016 likelihoods high-$\ell$ TT,TE,EE + LOWSim \cite{Aghanim:2016yuo}. \label{fig:fzxeAccretion}}
\end{figure}

\begin{figure}
\centering
\includegraphics[scale=0.38]{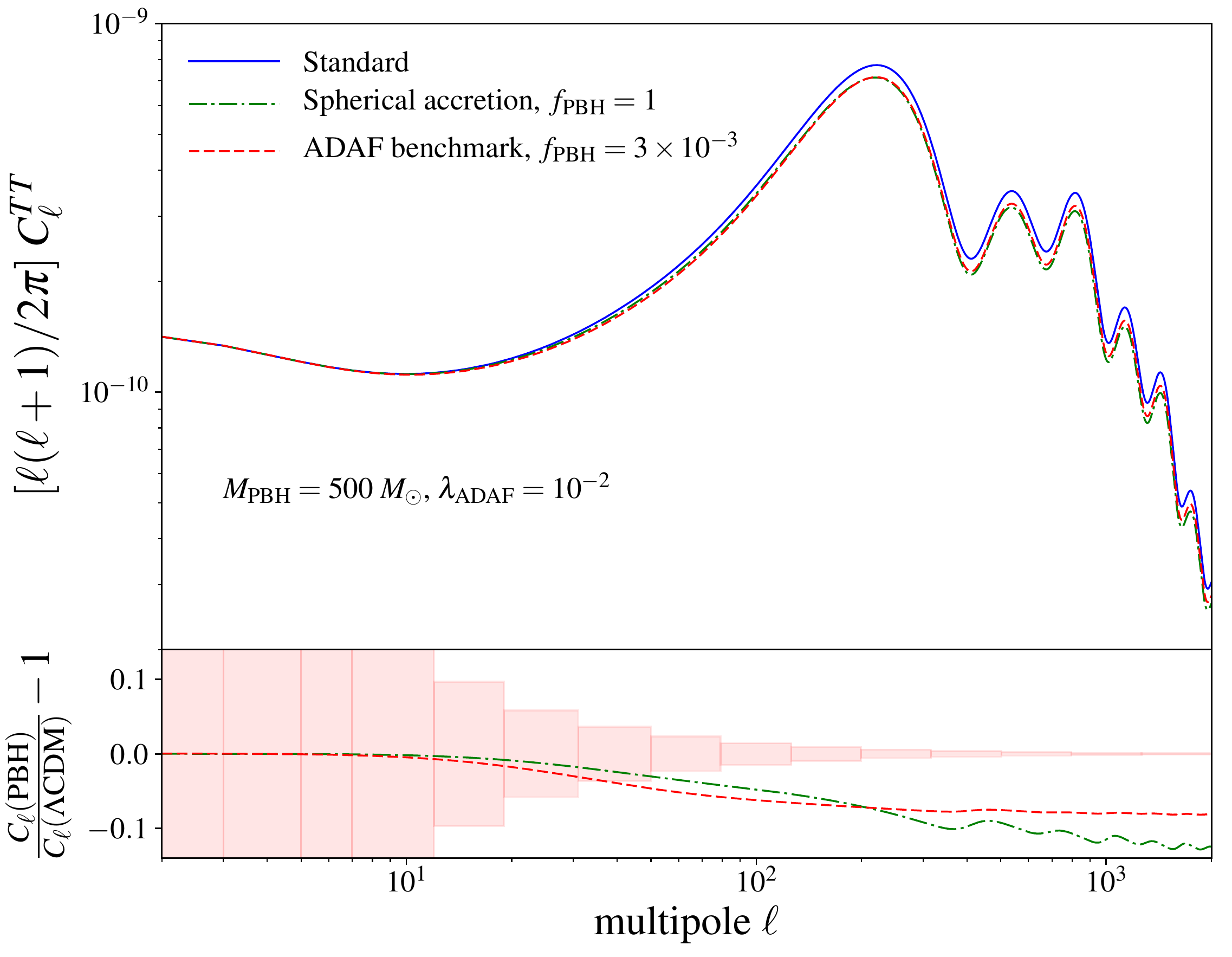}
\includegraphics[scale=0.38]{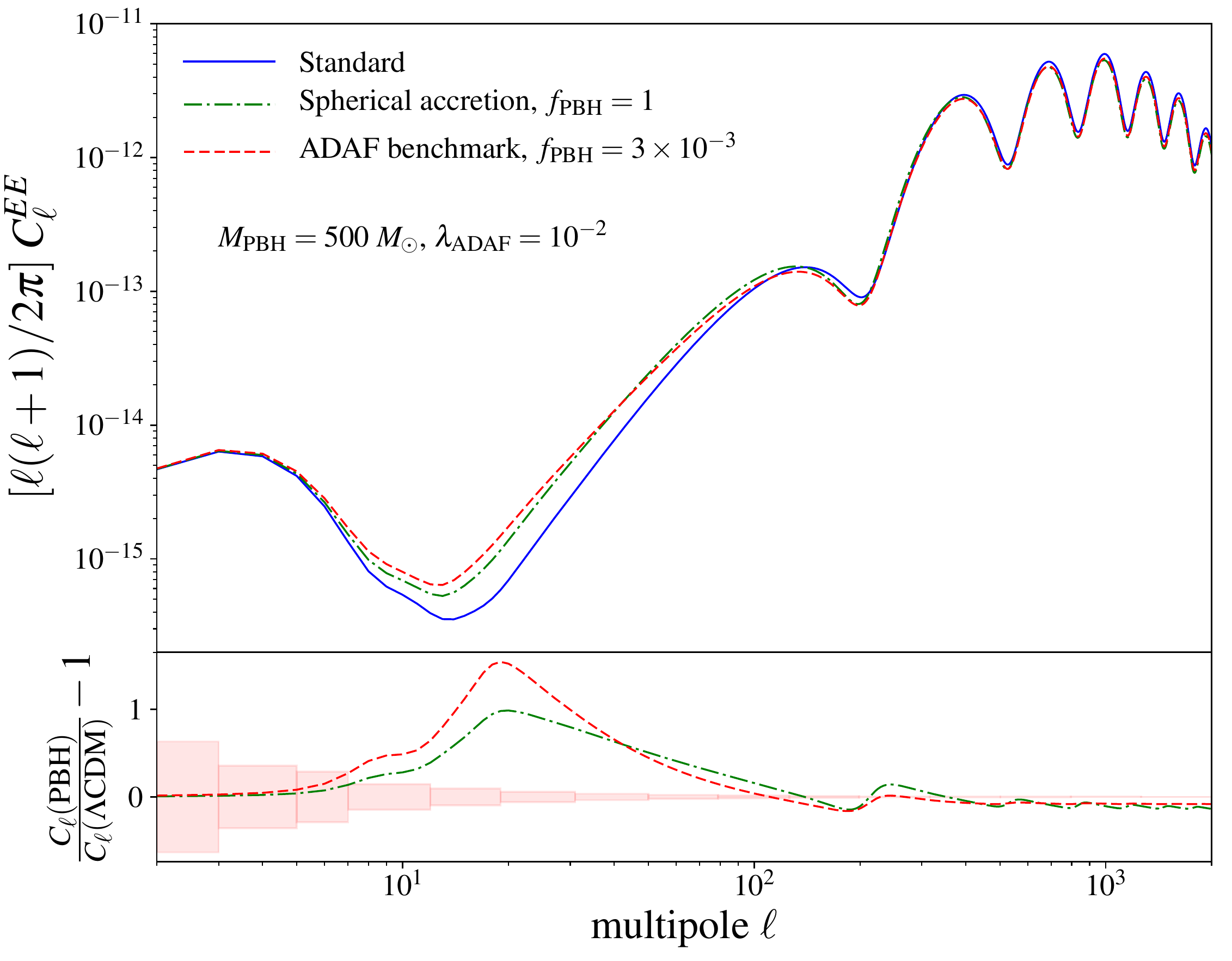}
\caption{CMB TT (top panel) and EE (bottom panel) power spectrum obtained for a monochromatic population of PBH with masses 500 $M_\odot$ depending on the accretion recipe used. \label{fig:Cl}}
\end{figure}

For consistency with our discussion in Sec.~\ref{Lumin}, we base our disk accretion spectra on the numerical results for ADAF models reported in Ref.~\cite{Yuan:2014gma}, Fig. 1. In particular, we adopt
\begin{equation}\label{eq:L_nudisk}
L_{\omega}\propto \Theta(\omega-\omega_{\rm min})\omega^{-a}\exp(-\omega/T_s)\,,
\end{equation}
with a choice for $T_s$ as above. We ignore the dependence of $T_s$ upon accretion rate and PBH mass, which is very mild in the range of concern for us. We consider $a\in[-1.3;-0.7]\,$, with a hardening linear in the log of $\dot{M}$ (as from the caption in that figure) with $-0.7$ corresponding almost to the limiting case of the thick disk.  We take  $\omega_{\rm min} =(10\,M_\odot/M)^{1/2}\,$eV. 
 Note that such cutoff at low energy only affects the normalization at the denominator of Eq.~(\ref{fzexpr}), i.e. the ``useful'' photon fraction of the bolometric luminosity, normalized as described in Sec.~\ref{Lumin}.  On the other hand, the cutoff at the numerator in Eq.~(\ref{fzexpr}) is in principle given by the ionization or excitation threshold (depending on the channel), since photons of lower energy do not contribute to the efficiency. In practice, the transfer functions are only directly available for energy injection above 5 keV. However, we can safely extrapolate the transfer function down to $\sim$ 100 eV: It has been shown in Ref.~\cite{Galli13} that the energy repartition fractions are to an extremely good approximation independent of the initial particle energy in the range between  $\sim$ 100 eV and a few keV. In fact, this behaviour is at the heart of the ``low energy code''  used by authors of Ref.~\cite{Slatyer15-2} to compute their transfer functions. Below  $\sim$ 100 eV, the power devoted to ionization starts to drop, and we conservatively cut the integral at the numerator at this energy.
We show the $f_c(z,x_{\rm e})$-functions for the spherical accretion scenario and the disk accretion scenario in Fig.~\ref{fig:fzxeAccretion} - top panel  (we chose a mass which we estimate to be among the least efficient at depositing energy). 
We incorporated the effects of accretion into a modified version  of the {\sf Recfast} module \cite{Seager:1999bc} of the Boltzmann solver {\sf CLASS} \cite{Blas:2011rf}.  It is enough for our purpose to work with a modified {\sf Recfast} that has been fudged to reproduce the more accurate calculation from {\sf CosmoRec} \cite{2011MNRAS.412..748C} and {\sf HyRec} \cite{2011PhRvD..83d3513A}.
The impact of the accretion on the free-electron fraction for a PBH mass of  $500 M_\odot$  is shown in the bottom panel of Fig.~\ref{fig:fzxeAccretion}: It is much more pronounced in the disk accretion scenario (we chose a PBH fraction $\sim$  300 times smaller!), even if the energy deposition efficiency is lower. 
In Fig.~\ref{fig:Cl}, the corresponding impact on the CMB power spectra is illustrated. The effects are typical of an electromagnetic energy injection (for a detailed review see Ref.~\cite{2017JCAP...03..043P}): The delayed recombination slightly shifts acoustic peaks and thus generates small wiggles at high multipoles $\ell$ in the residuals with respect to a standard $\Lambda$CDM scenario. Meanwhile, the increased freeze-out fraction leads to additional Thomson scattering of photons off free electrons along the line-of-sight, which manifests itself as a damping of temperature anisotropies and an enhanced power in the polarization spectrum. Note that in principle the different accretion recipes could be distinguished via a CMB anisotropy analysis. Indeed, each accretion scenario has a peculiar energy injection history which does not lead to a simple difference in the normalization: the actual shape of the power spectra slightly changes. This behavior is also present when changing the PBH mass, but is much less pronounced, albeit still above cosmic variance in the EE spectrum (not shown  here to avoid cluttering). Hence, if a signal were found, it is conceivable that some constraints could be put on the PBH mass  and (especially) accretion mechanism, but a strong statement would require better characterization of the signal, which goes beyond our present goals.

We compute the 95\% CL bounds using data from {\em Planck} high-$\ell$ TT TE EE+lensing  \cite{Planck15} and a prior on $\tau_{\rm reio}$ \cite{Aghanim:2016yuo}, by running an MCMC using the {\sf MontePython} package \cite{Audren12} associated to {\sf CLASS}.  For ten PBH masses log-spaced in the range $[M_{\rm min},1000 M_\odot]$ we perform a fit to the data with flat priors on the following set of parameters:
$$
\Lambda{\rm CDM}\equiv\{\omega_b,\theta_s,A_s,n_s,\tau_{\rm reio},\omega_{\rm DM}\}+f_{\rm PBH}\,,
$$
with $M_{\rm min}$ fixed by a preliminary run where $f_{\rm PBH}$ has been set to one, and the PBH mass $M_{\rm PBH}$ has been let free to vary (with a flat prior as well)\footnote{We have checked that making use of a logarithmic prior {\it improves} the bound by roughly $50\%$. We thus conservatively stick to the linear prior, which also eases comparison to previous works.}.
We use a Choleski decomposition to handle the large number of nuisance parameters in the Planck likelihood \cite{Lewis:2013hha}. We consider chains to have converged when the Gelman-Rubin \cite{Gelman:1992zz} criterium gives $R -1<0.01$. 
First, to check our code, we run it {\it under the same hypotheses as}~\cite{Ali-Haimoud:2016mbv}
(the conservative, collisional ionization case), finding the constraint $M_{\rm PBH}< 150\,M_\odot$ for $f_{\rm PBH}=1$, as opposed to their $M_{\rm PBH}\lesssim 100\,M_\odot$. We attribute the 50\% degradation of our bound compared to Ref.~\cite{Ali-Haimoud:2016mbv} to our more refined energy deposition treatment. We checked that an agreement at a similar level with Refs.~\cite{Ricotti:2007jk,Horowitz:2016lib} is obtained if we implement their prescriptions, but since some equations in Ref.~\cite{Ricotti:2007jk} (re-used in Ref.~\cite{Horowitz:2016lib}) have been shown to be erroneous~\cite{Ali-Haimoud:2016mbv}, we do not discuss them further.

\begin{figure}[!t]
\hspace{-2.3cm}
\vspace{-1.1cm}
\includegraphics[scale=0.30]{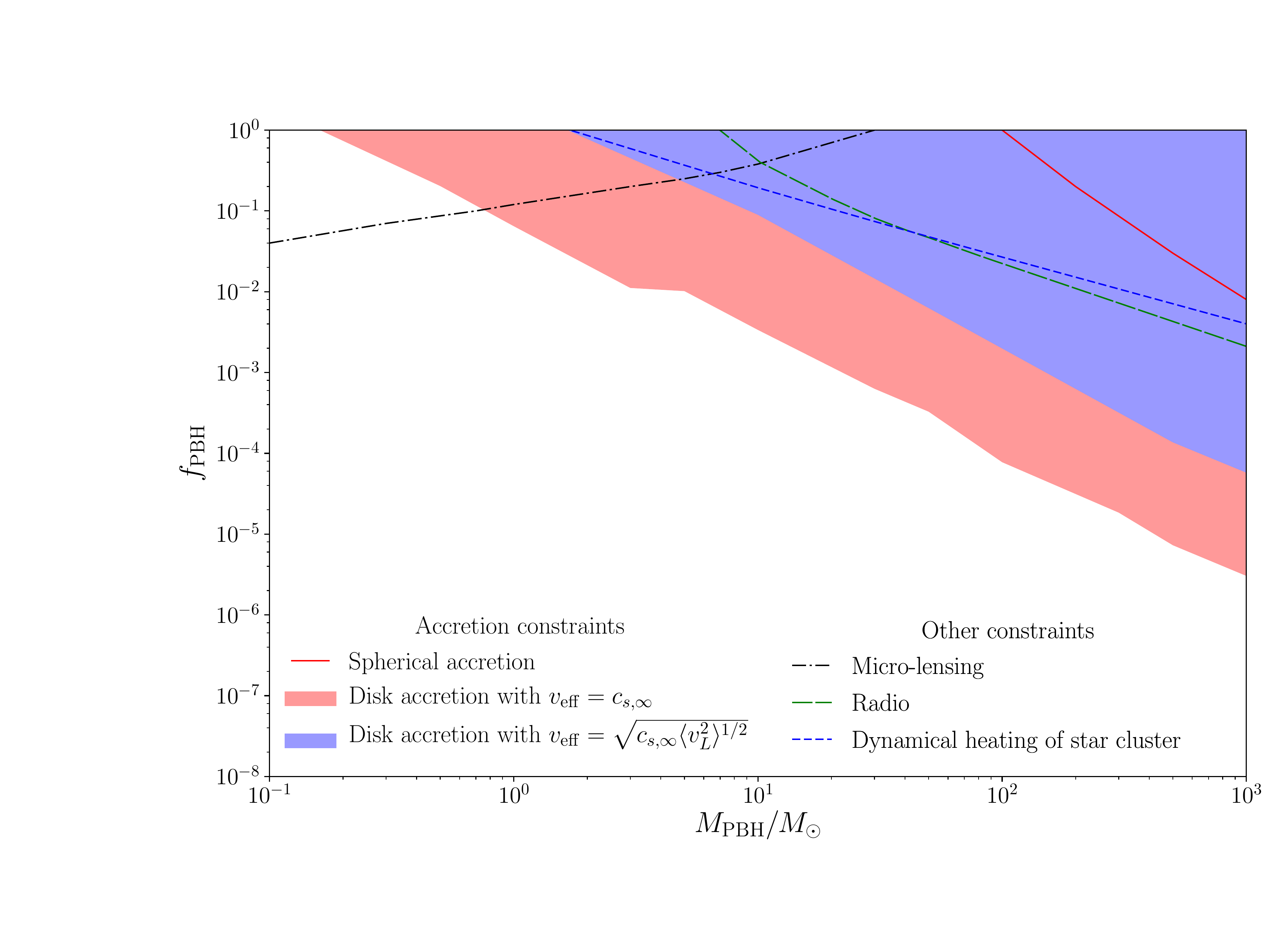}
\caption{ Constraints on accreting PBH as DM. Our constraints, derived from a disk accretion history (blue region: Eq.~(\ref{veff}); light-red region: $v_{\rm eff}\simeq c_{s,\infty}$), are compared to: i) the CMB constraints obtained assuming that spherical accretion holds as in Ref.~\cite{Ali-Haimoud:2016mbv} (red full line); ii) the non observation of micro-lensing events in the Large Magellanic Cloud as derived by the EROS-2 collaboration \cite{Tisserand:2006zx} (black dot-dashed line); iii) the non observation of disk-accreting PBH at the Galactic Center in the radio band, extrapolated from Ref.~\cite{Gaggero:2016dpq} (green long-dashed line); iv) constraints from the disruption of the star cluster in Eridanus II \cite{Green:2016xgy} (blue short-dashed line, see text for details).\label{fig:monochromatic}}
\end{figure}

Our fiducial conservative constraints (at 95\% C.L.) are represented  in Fig.~\ref{fig:monochromatic} with the blue-shaded region  in the plane $(M_{\rm PBH},f_{\rm PBH})$: We exclude PBH with masses above $\sim 2\, M_\odot$ as the dominant form of DM. The constraints can be roughly cast in the form:
\begin{equation}
f_{\rm PBH} < \bigg(\frac{2\,M_\odot}{M}\bigg)^{1.6}\bigg(\frac{0.01}{\lambda}\bigg)^{1.6}\,.\label{fidbound}
\end{equation}
This is two orders of magnitudes better than the spherical accretion scenario, and it improves significantly over the radio and X-ray constraints from Ref.~\cite{Gaggero:2016dpq}, without dependence on the DM halo profile as those ones. 
Lensing constraints are nominally better only at  $M\lesssim 6\,M_\odot$. Note also the importance of the relative velocity between PBH and accreting baryons: If instead of Eq.~(\ref{veff}) we were to adopt $v_{\rm eff}\simeq c_{s,\infty}$---representative of a case where a density of baryons comparable to the cosmological one is captured by halos at high redshift---the bound would improve by a further order of magnitude, to $M\lesssim 0.2\, M_\odot$ (light-red shaded region in Fig. \ref{fig:monochromatic}). This is also true, by the way, for the spherical accretion scenario, where---all other conditions being the same---adopting $v_{\rm eff}\simeq c_{s, \infty}$ would imply $M\lesssim 15\,M_\odot$, to be compared to $M\lesssim 150 M_\odot$ previously quoted. The ``known'' uncertainties in disk accretion physics are probably smaller: When varying---at fixed accretion eigenvalue $\lambda$---the electrons heating parameter $\delta$ within the range described in section \ref{sec:disks}, for the 30 $M_\odot$ benchmark case reported in the bottom panel of Fig.~\ref{fig:PBHAccretion}, the radiative efficiency $\epsilon$ varies by a factor $\sim 3$, reflecting correspondingly on the constraints. To help the readers grasp the dependence of the bound upon different parameters, 
we also derive a parametric bound, obtained from a run where we assumed that $v_{\rm eff}$ is constant over time (and the accretion rate is always small, i.e. $\dot{M}_{\rm B} < 10^{-3}L_{\rm Ed}$), scaling as
\begin{equation}
f_{\rm PBH} < \bigg(\frac{4\,M_\odot}{M}\bigg)^{1.6}\bigg(\frac{v_{\rm eff}}{10~{\rm km/s}}\bigg)^{4.8}\bigg(\frac{0.01}{\lambda}\bigg)^{1.6}\,.\label{parambound}
\end{equation}
\vspace{-0.9cm}
\vspace{+0.4cm}
\begin{figure}[!hb]
\hspace{-0.5cm}
\includegraphics[scale=0.29]{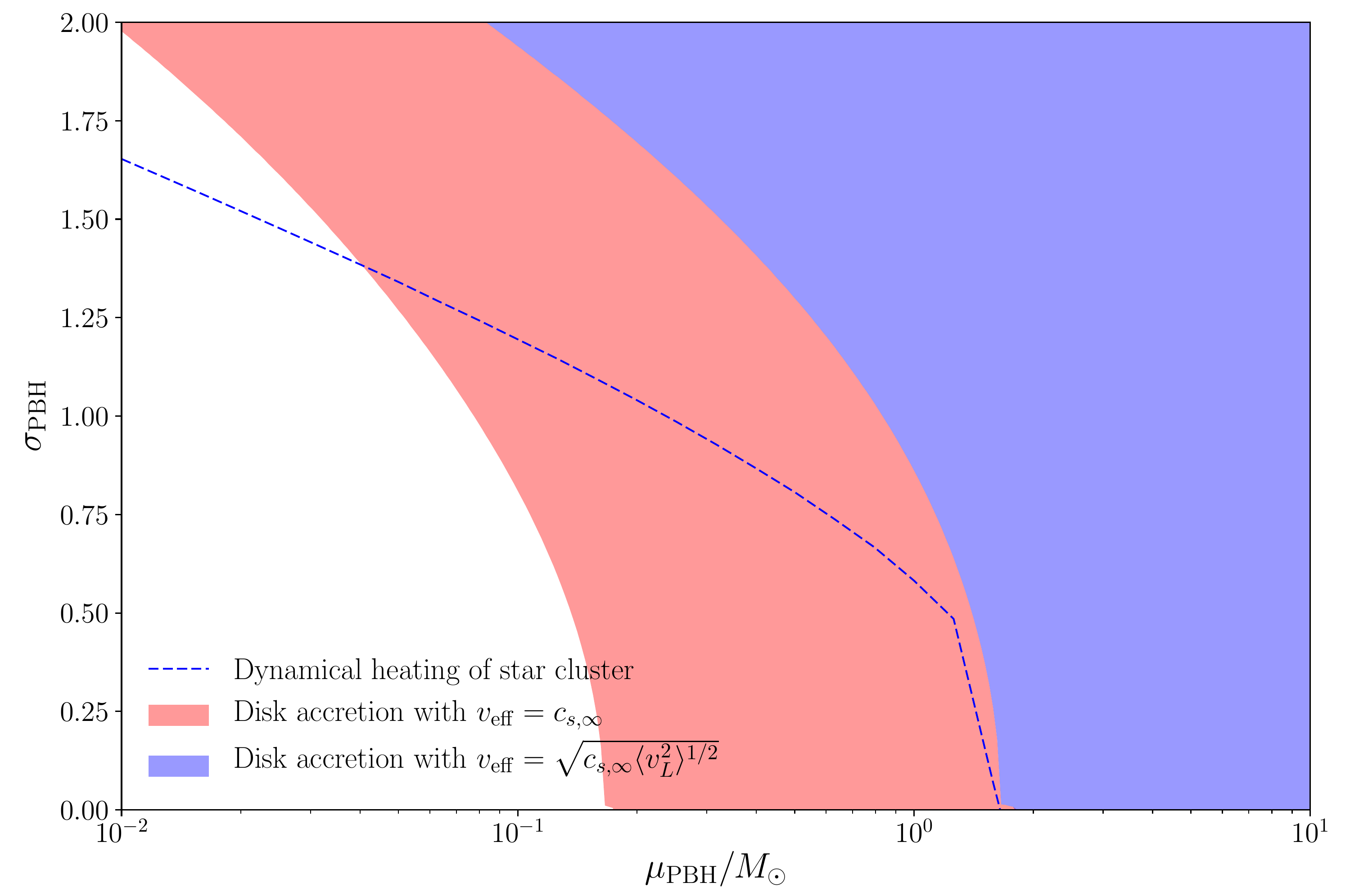}
\caption{Constraints on the width $\sigma_{\rm pbh}$ of a broad mass spectrum of accreting PBH  as from Eq.~(\ref{eq:broad_spectrum}) as a function of the mean mass $ \mu_{\rm PBH}$, assuming that they represent 100\% of the DM.  For comparison the dashed blue line represents our calculation of the best constraint from the dynamical heating of the star cluster in the faint dwarf Eridanus II, following the method and parameters of Ref.~\cite{Green:2016xgy}. }\label{fig:broad_mass}
\end{figure}

We have also extended the constraints to a broad log-normal mass distribution of the type 
\begin{equation}\label{eq:broad_spectrum}
M\frac{\dd n}{\dd M}=\frac{1}{\sqrt{2\pi}\sigma M}\exp\bigg(\frac{-\log_{10}(M/\mu_{\rm PBH})^2}{2\sigma_{\rm pbh}^2}\bigg)\,.
\end{equation}
i.e. with mean mass $\mu_{\rm PBH}$ and width  $\sigma_{\rm pbh}$. Our constraints in the plane $(\sigma_{\rm pbh}, \mu_{\rm PBH})$  assuming that PBH represent 100\% of the DM are shown in Fig.~\ref{fig:broad_mass}. It is clear that the bound on the median PBH mass is robust and can only get more stringent if a broad, log-normal mass function is considered, confirming the overall trend discussed in Ref.~\cite{Carr:2017jsz}. However, we estimate that the tightening of the constraints for a broad mass function is more modest than the corresponding one from some dynamical probes. This is illustrated by the blue dashed line in Fig.~\ref{fig:broad_mass}, which 
is the result of our calculation of the constraints from the disruption of the star cluster in Eridanus II, 
following the method and parameters of Ref.~\cite{Green:2016xgy} (cluster mass of $3000 \ M_\odot$, timescale of $12$ Gyr, initial and final radius of $2$ pc and $13$ pc respectively and a cored DM density of $\rho_{\rm DM} = 1 M_\odot {\rm pc}^{-3}$).

\section{Conclusions}\label{sec:conclu}

The intriguing possibility that DM is made of PBH is nowadays a subject of intense work in light of the recent gravitational wave detections of merging BH with masses of tens of $M_\odot$. However, high mass PBH are known to accrete matter, a process that leads to the emission of a high energy radiation able to perturb the thermal and ionization history of the universe, eventually jeopardizing the success of CMB anisotropy studies. In this computation, the geometry of the accretion, namely whether it is spherical or associated to the formation of a disk, is a major ingredient. Until now, studies have focused on the case of spherical accretion. In this work, we argued that, based on a standard criterion for disk formation, all plausible estimates suggest that a disk forms {\em soon after recombination}. This is essentially due to the fact that  stellar-mass PBH are in a non-linear regime (i.e. clustered in halos of bound objects, from binaries to clumps of thousands of PBH) at scales encompassing the Bondi radius already {\em before recombination}. This feature was ignored in the pioneering article~\cite{Ricotti:2007au}, which assumed that massive PBH cluster like WIMPs and deduced the adequacy of the spherical accretion approximation, eventually adopted by all subsequent studies.

Then, we have computed the effects of accretion around PBH onto the CMB power spectra, making use of state-of-the art tools  to deal with energy deposition in the primordial gas. Our 95\% CL fiducial bounds  preclude PBH from accounting for the totality of DM if having a monochromatic distribution of masses above $\sim 2\, M_\odot$, the bound on $f_{\rm PBH}$ improving roughly like $M^{1.6}$ with the mass. All in all, the formation of disks improves over the spherical approximation of Ref.~\cite{Ali-Haimoud:2016mbv} by two orders of magnitude. We also checked that the constraints derived on the monochromatic mass function apply to the average mass value of a broad,  log-normal mass distribution too, actually becoming more stringent if the distribution is broader than a decade. 

A realistic assessment of ``known'' astrophysical uncertainties, like for instance the electron share of the energy in ADAF models, suggests that our quantitative results can only vary within a factor of a few, not enough to change qualitatively our conclusions. Nonetheless, we believe that our constraints are conservative rather than optimistic. In particular, we assumed accretion from an environment at the {\it average cosmological density}: This is less and less true when PBH halos gradually capture baryonic gas in their potential wells. Alone, capturing from a pool of baryons of density comparable to the cosmological one, but bound to PBH halos, would reduce the relative PBH-baryon velocity and improve the bounds to $\sim 0.2 M_\odot$. Once baryons accumulate well above the cosmological average, the accretion rate $\dot{M}$ from this bound component grows correspondingly, and the constraining power more than linearly with it. It would be interesting to reconsider the CMB bounds on stellar-mass PBH once a better understanding of the halo assembly history in these scenario is achieved, a task probably requiring dedicated hydrodynamical simulations. 

Together with other constraints discussed recently (see for instance~\cite{Koushiappas:2017chw,Brandt:2016aco,2014ApJ...790..159M,Green:2016xgy,Gaggero:2016dpq,Inoue:2017csr}) our bounds suggest that the possibility that PBH of stellar masses could account for an appreciable fraction of the DM is excluded. It remains to be seen if the small $f_{\rm PBH}$ allowed by present constraints may still be sufficient to explain LIGO observations in terms of PBH and, in that case, to find signatures of their primordial nature, possibly peculiar of some specific production mechanism: Such signatures become all the more crucial since both PBH mass (of stellar size) and their small DM fraction (for instance, in a halo of the Milky Way size about 0.1\% of the DM should be made of astrophysical BH) cannot be easily used as diagnostic tools to discriminate PBH from astrophysical ones. 
 It is worth noting that, based on the recent study \cite{2017JCAP...03..043P}, we expect that forthcoming CMB polarization experiments (very sensitive to energy injection) and 21 cm experiments~\cite{2013MNRAS.435.3001T,Gong:2017sie} (the golden channel for searches looking at energy-injection during the Dark Ages) will be able to give more insights on PBH scenarios, including stellar mass ones, even if the possibility that they may contribute to a high fraction of the DM has faded away.

\begin{acknowledgments}
This work is partly supported by the Alexander von Humboldt Foundation (P.S.), JSPS KAKENHI Grant Numbers 26247042, 
JP15H05889, JP16H0877,  JP17H01131 (K.K.), the Toshiko Yuasa France-Japan Particle Physics Laboratory ``TYL-FJPPL'' (P.S. and K.K.),
as well as ``Investissements d' avenir, Labex ENIGMASS'' of the French ANR (V.P.). The authors warmly thank Yacine Ali-Hamoud, Juan Garc\`ia-Bellido, Mark Kamionkowski, Nagisa Hiroshima, and Ville Vaskonen for useful comments and discussions, and J. Lesgourgues for discussions and technical help with the {\sf CLASS} implementation. 
\end{acknowledgments}

\bibliography{biblio}
\end{document}